\documentclass{aa} 
\usepackage{natbib, my_def, graphicx} 
\bibpunct{(}{)}{;}{a}{}{,}
\bibstyle{aa-6.0}     
\def\udier{\"u}

\def \etal   {{~et~al.~}}
\begin{document}
\title{Radio polarimetry of compact steep spectrum sources at sub-arcsecond resolution}
 
\author{ F. Mantovani \inst{1,2} \and
         A. Rossetti  \inst{1} \and
	 W. Junor     \inst{3} \and
	 D.J. Saikia  \inst{4,5} \and
	 C.J. Salter  \inst{6} 
         }

\offprints{Franco Mantovani\\
  \email{fmantovani@ira.inaf.it}}

\institute{Istituto di Radioastronomia -- INAF, Via Gobetti 101,
 I--40129, Bologna, Italy
 \and Max-Planck-Institut f\"{u}r Radioastronomie, Auf dem H\"{u}gel 69, D-53121 Bonn, Germany 
 \and Los Alamos National Laboratory, Los Alamos, NM 87545, USA
 \and Cotton College State University, Panbazar, Guwahati 781 001, India
 \and National Centre for Radio Astrophysics, TIFR, Post Bag 3,
 Ganeshkhind, Pune 411 007, India
 \and Arecibo Observatory, HC3 Box 53995, Arecibo, Puerto Rico 00612
}

\date{Received \today; accepted ???}

\abstract
{}
{We report new Very Large Array polarimetric observations of Compact
Steep-Spectrum (CSS) sources at 8.4, 15, and 23\,GHz. }
{Using multi-frequency VLA observations we have
derived sub-arcsecond resolution images of the total intensity,
polarisation, and rotation measure (RM) distributions.
}
{  We present multi-frequency VLA polarisation observations of
CSS sources.
About half of the sources are point-like even at the resolution
of $\sim0.1\times0.1$ arcseconds. The remaining sources have double or
triple structure.
Low values for the percentage of polarised emission in CSS sources
is confirmed. On the average, quasars are more polarised than galaxies. 
A wide range of RM values have been measured. There are clear indications 
of very large RMs up to $\approx$5\,585 rad m$^{-2}$. 
CSS galaxies are characterized by RM values that are larger than CSS quasars.
The majority of the objects show very large values of RM.
}
{The available data on sub-arcsecond-scale rest-frame RM estimates for 
CSS sources show that these have a wide range of values extending up
to $\sim$36\,000 rad m$^{-2}$. RM estimates indicate an overall density
of the magneto-ionic medium larger than classical radio sources.
}

\keywords{polarisation -- galaxies: quasars: general -- galaxies: jets -- 
          radio continuum: galaxies}

\maketitle 
\section{Introduction} 
The number of Compact Steep-Spectrum (CSS) sources with detailed
polarimetric information available at sub-arcsecond resolution is
still small.   We have conducted a series of polarimetric observations of
CSS sources using the Very Large Array (VLA).

CSS objects are {\it young} radio sources, with ages $<10^{{\rm 3-5}}$\,yr
 \citep {Fanti90}.
They have linear sizes $\leq 20$\,kpc
\footnote {$H_0=71\,{\rm km}\, {\rm s}^{-1}\, {\rm Mpc}^{-1}, \Omega_{\rm m}=0.27, \Omega_{\rm vac}=0.73$} 
and steep high-frequency radio spectra ($\alpha >0.5$; ${\rm
S}_{\nu}\propto\nu^{-\alpha}$). Being sub-galactic in size, CSS sources
reside within their host galaxies.  Therefore, Faraday rotation
effects are to be expected when their polarised synchrotron emission is
observed through the magneto-ionic interstellar medium (ISM) of the
host galaxy. The comparison of polarised emission over a range of
wavelengths is an important diagnostic of the physical conditions
within and around these compact radio sources.

Existing sub-arcsec polarimetry has provided evidence in favour of
the interaction of components of CSSs with dense gas clouds, \citep
[for example, see] [] {Junor99a}.

Aiming at a deeper understanding of the CSS source phenomenon, 
with the VLA 
\footnote {The National Radio Astronomy Observatory is a facility of the 
National Science Foundation operated under cooperative agreement by 
Associated Universities, Inc.} 
A-Array we observed an ``incomplete'' 
sample of 29 sources selected from the list of \citet{Dallacasa90}. 
The adopted selection criteria were: i) total flux
density at 5\,GHz $>$1\,Jy, ii) declination $>-20^\circ$,
and iii) lack of observations at sub-arcsecond resolution (at the time of 
source selection).

In this paper, we report multi-frequency VLA polarisation observations 
of our CSS sample at 8.1, 8.5, 15 and 23\,GHz.

In Section \ref{sec:observation} we summarise the observations and data
processing. Section \ref{sec:results} describes the new information
obtained on the structural and polarisation properties.
Discussion and conclusions are
presented in Sections \ref{sec:discussion}  and \ref{sec:summary}
respectively.

\section{Observations and data reduction} 
\label{sec:observation}

Polarimetric observations of our sample of 29 CSS sources were made
at 8, 15 and 23\,GHz on August 6th, 1991 using the VLA in
A-array.  The observations were scheduled in ``bracket'' mode, i.e.
calibrator$-$source$-$calibrator, to obtain the best possible phase
correction.  Despite this, almost half of the data acquired at 23\,GHz
could not be imaged due to poor weather conditions. The data were
recorded in both circular polarisations, and calibrated in the
standard way using AIPS procedures.  The sources 2200$+$420 (BL~Lac) and
0923$+$392 were observed regularly throughout the observing period to
allow for parallactic angle corrections.  Calibration of the Electric
Vector Position Angle (EVPA) was performed by observing the source
1328$+$307 (3C286) assuming an EVPA of $33^\circ$ for it at all
frequencies.  An iterative procedure was carried out using the IMAGR
and CALIB programmes to self-calibrate the parallel-hand (LL or RR)
fringes. The complex gain corrections  derived in this way were
also applied to the cross-hand (RL or LR) fringes. Images in Stokes
parameters I, Q, and U were produced. Images of the polarised flux
density P=(Q$^2+$U$^2$)$^{0.5}$ and EVPA, $\chi = 0.5\times$tan$^{-1}$(U/Q), 
were then generated from the Q and U images.
The data acquired with the 8\,GHz receiver were separately imaged for
the two IFs, namely IF1 at 8085\,MHz and IF2 at 8485\,MHz,  each
having a bandwidth of 50\,MHz. When considered together, observations 
at these two frequencies are designated ``X-band'' in the text.  

\section{Results} 
\label{sec:results} 
\subsection{Images}

The full-resolution images are available on the web site
http://db.ira.inaf.it/aj206-fm/.

In practice, half of the sources (15) are point-like at our 
resolution. However, two of 
these are slightly extended on one side. Seven others show double 
structure, and six more are triples. One further source is 
classified as complex.
Examples of sources classified as above are presented in 
Fig.~\ref{fig:structure}. This morphological description follows
that of \citep {Spencer91}.  The vectors representing  the electric 
vector of 
linearly polarised intensity are overlaid on the total-intensity contours.
\vspace{-0.3cm}
\begin{figure*}[t]
\addtocounter{figure}{+0}
\centering
\includegraphics[width=8cm]{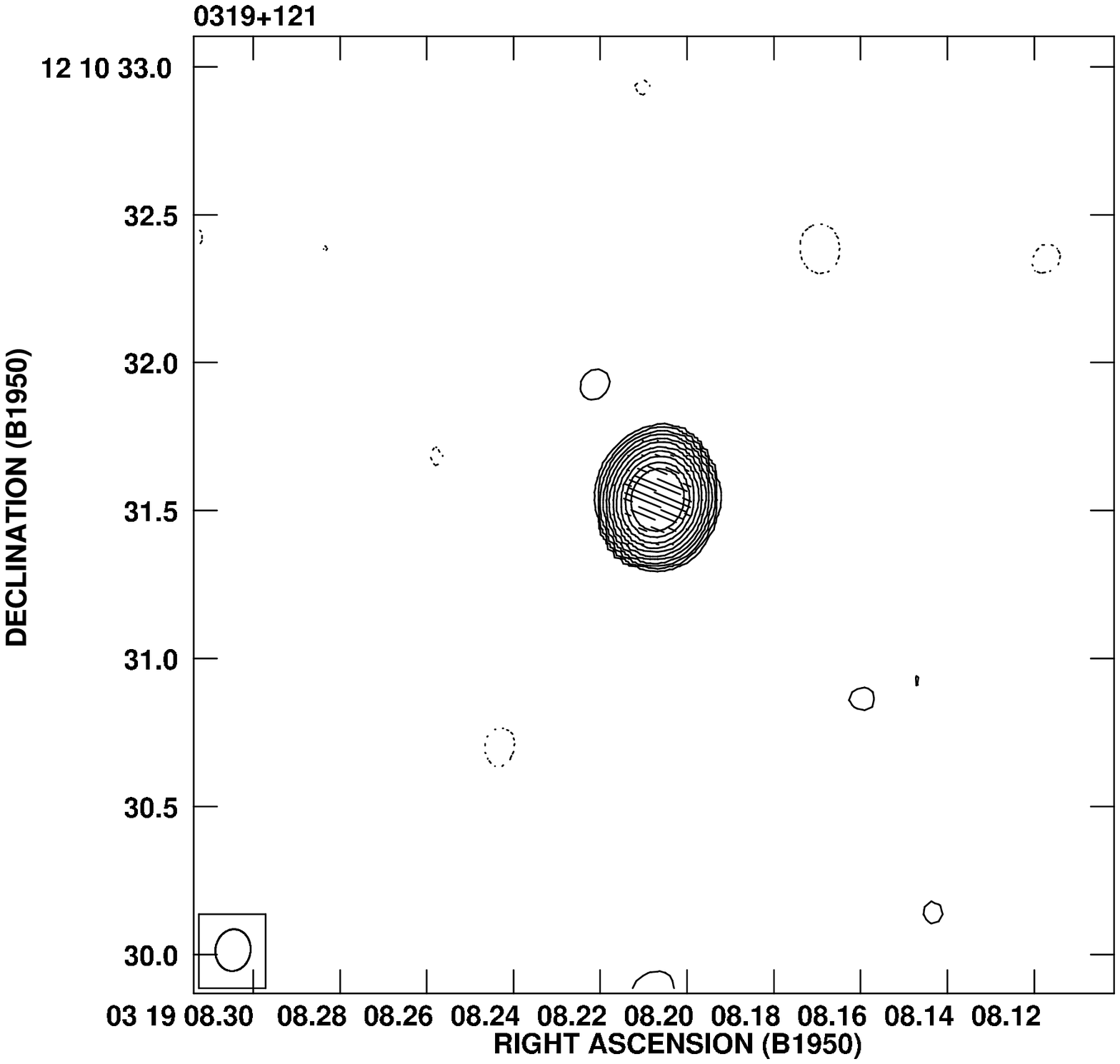}
\includegraphics[width=8cm]{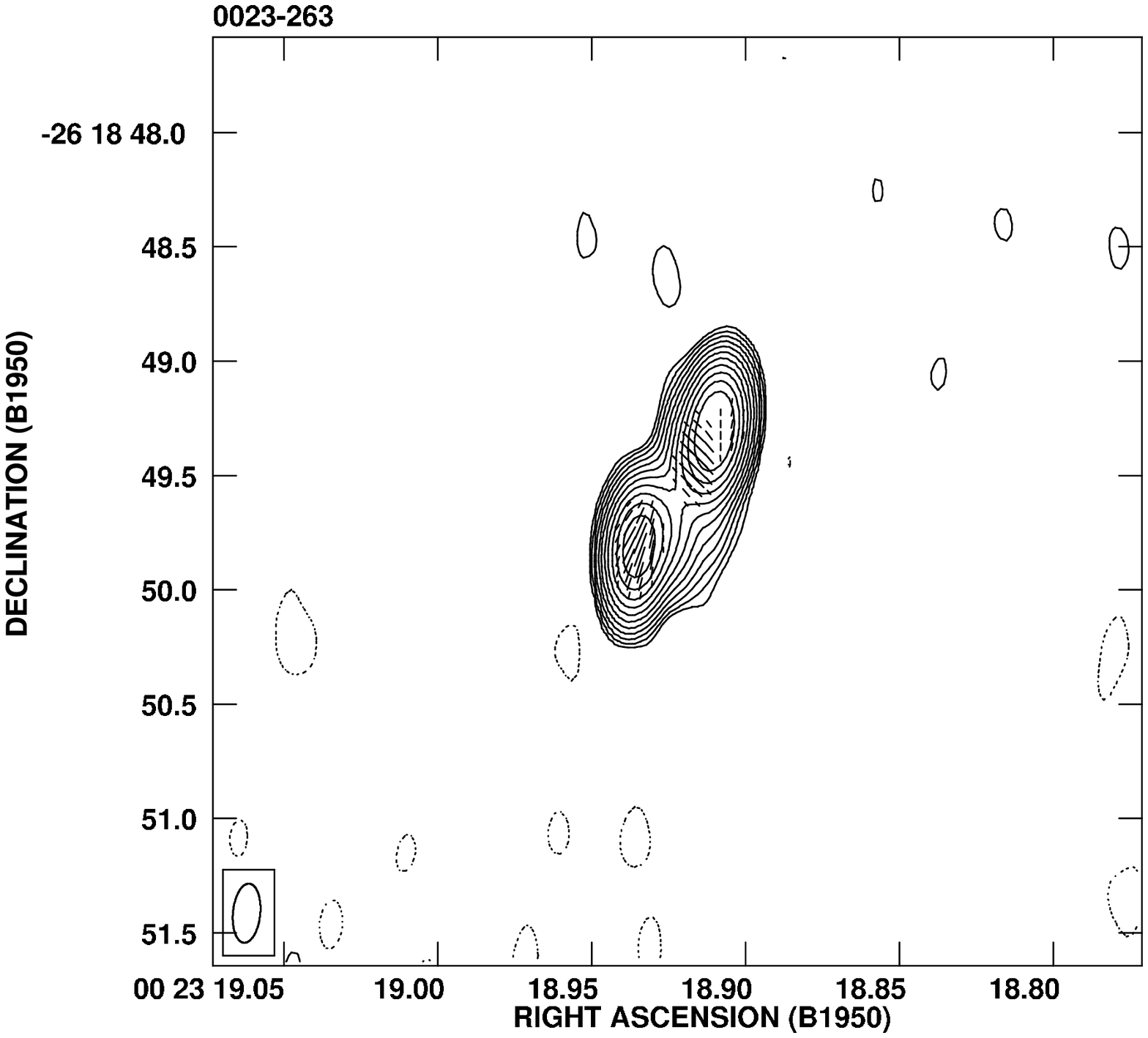}
\includegraphics[width=8cm]{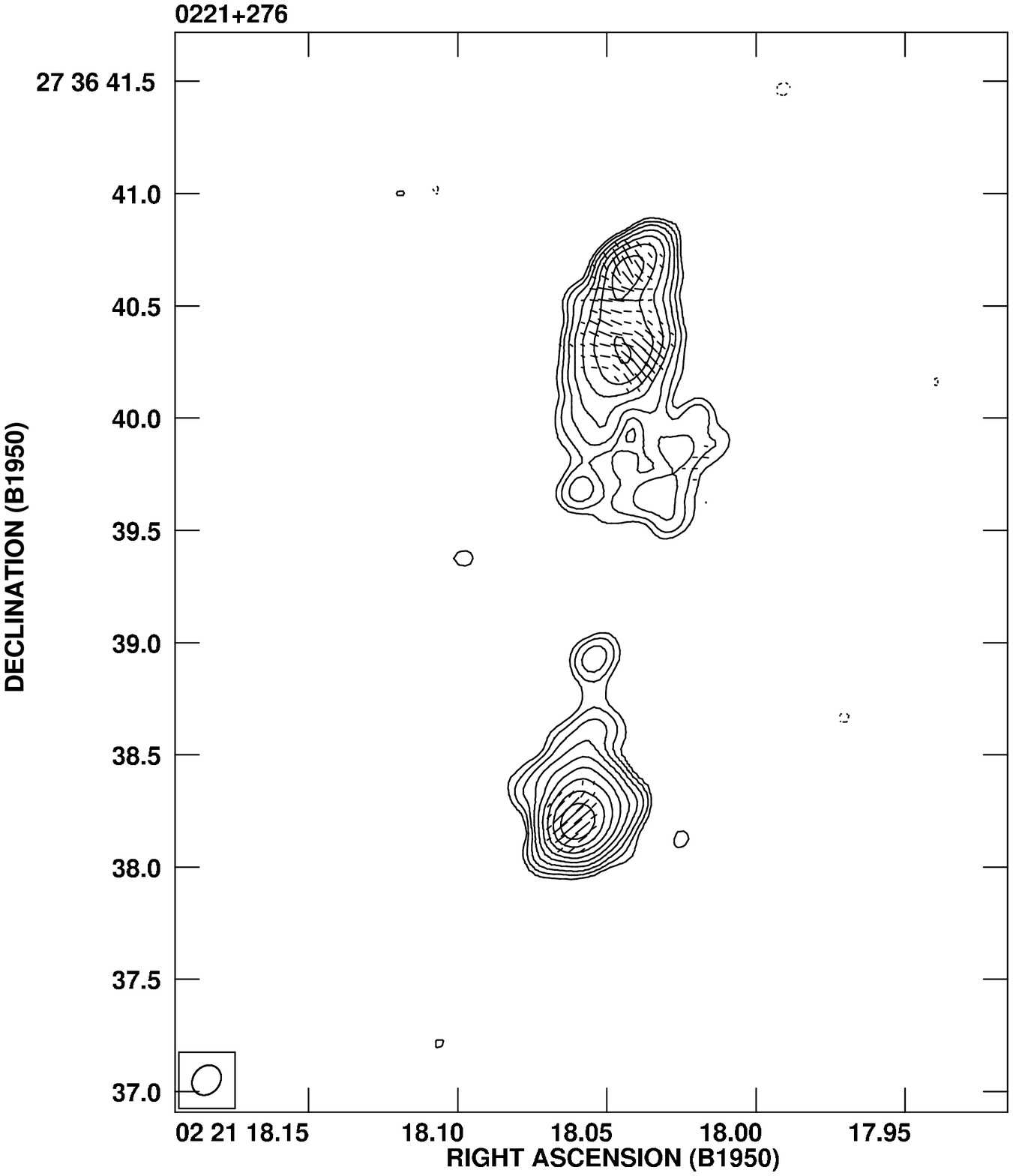}
\includegraphics[width=8cm]{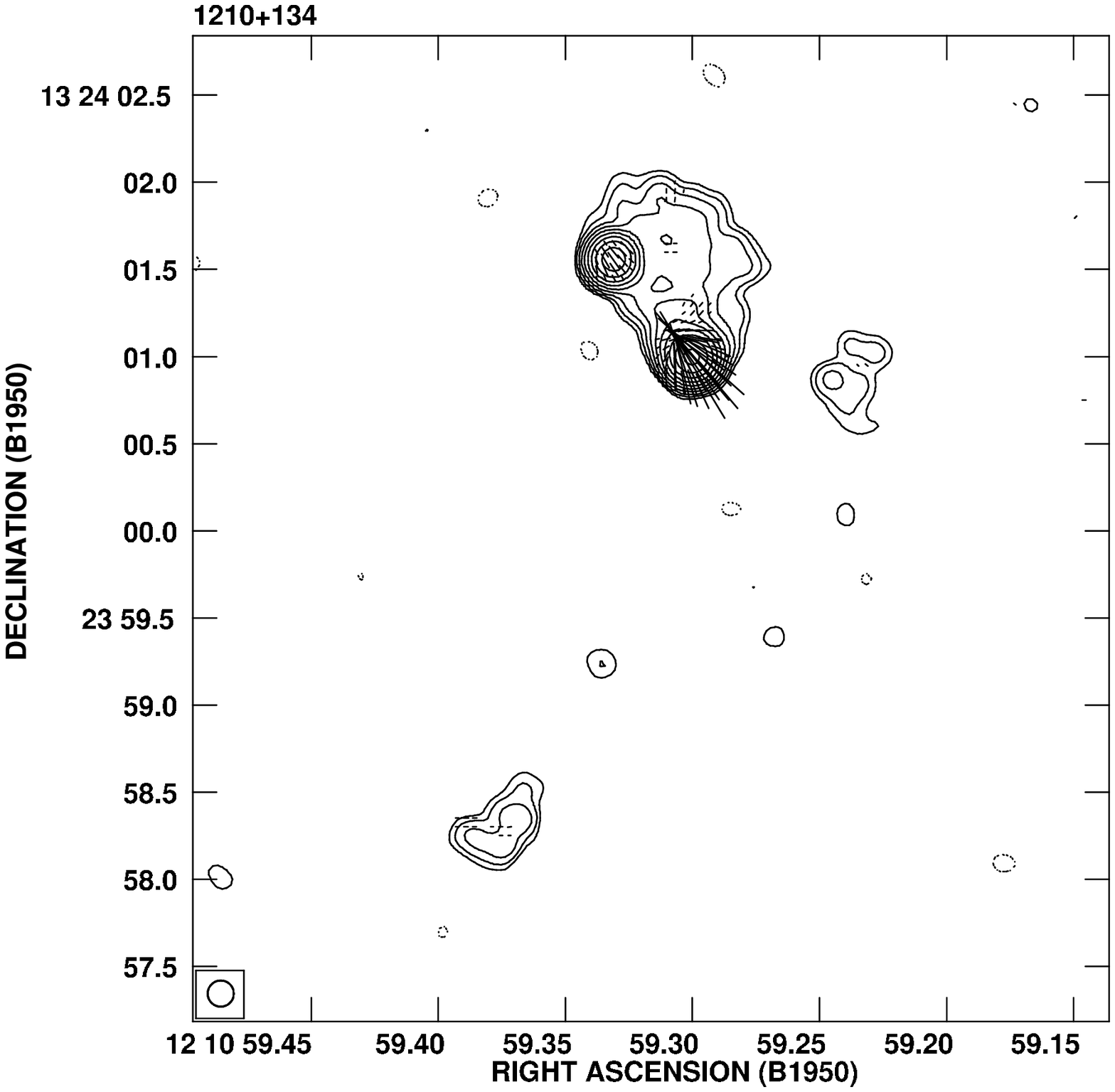}
\caption{The total-intensity images of a point-like source
(0319$+$121), a double source (0023$-$263), a triple source (0221$+$276), and
a complex source (1210$+$134) at 14885\,MHZ. The peak flux density is 
926.8\,mJy/beam, 635.8\,mJy/beam, 96.8\,mJy/beam, 161.3\,mJy/beam,
respectively. Contours are $-$1, 1, 2, 4, 8, 16, 32, 64,128, 256, 512, 1024
mJy/beam. An electric vector length of 1 arcsecond = 200\,mJy/beam for 
0319$+$121, and 20\,mJy/beam for others. 
\label{fig:structure}}
\end{figure*}
\subsection{Polarised emission}

The great majority of the sample sources show evidence of polarised
emission. However, generally, they are only weakly polarised. 
Three sources, namely 0531$+$194, 0941$-$080, and 1311$+$678, show no
polarised emission above the detection limits at any of the frequencies 
observed. The median percentage polarisations measured are summarized in 
Table~\ref{tab:summary-m}. The polarisation percentage integrated
over complete sources decreases from high to low frequencies, with
quasars more polarised than galaxies. It could be noted that there
are sources in which the polarisation percentage of individual components
are higher than the polarisation percentage integrated over the complete
source. This is due to averaging with a range of EVPAs .
\tabcolsep0.1cm
\begin{table*}[h]
\centering
\caption{ Percentage polarizations for the whole sample and for galaxies and quasars
separately}
\label{tab:summary-m}
\begin{tabular}{lccc}
\hline
Frequency        &  &$\langle m_{tot} \rangle$  \\
MHz              &  all      &    G          &   Q           \\
\hline
8085              & 1.1       &  0.5          &  2.8         \\
8485              & 1.1       &  0.3          &  2.8         \\
14885             & 3.0       &  1.1          &  4.2         \\
23285             & $\sim$3.5 &  $\sim$2.7    &  $\sim$4.0    \\
\hline
\end{tabular}
\normalfont
\smallskip\noindent
\flushleft{\normalsize {
The table is organised as follows:
Column 1: observing frequency; 
2-4: the values of the median percentage polarisation 
     $\langle m_{tot} \rangle$ (the ratio of polarised flux density 
     to the total flux density for the whole source) for the sample divided 
     into galaxies (G), and quasars (Q);
5: the median percentage polarisation for the source components considered 
   individually $\langle m_{comp} \rangle$.
}}
\end{table*}

The classification of source structures, together with  percentage 
polarised emissions derived by integrating the total-intensity and the 
polarised-intensity emissions over the total extent of each source, and 
the depolarisation indices DP, defined as the ratio of the percentages 
of polarised emission at the lower to the higher frequencies, are reported 
in Table~\ref{tab:percentage}.
\tabcolsep0.1cm
\begin{table*}[h]
\centering
\caption{Integrated polarisation percentages for each source}
\label{tab:percentage}
\begin{tabular}{llcllrrrrrrr}
\hline
Name    & Other      & ID & z & rs &$m_{8.0}$ & $m_{8.4}$ & $m_{14.8}$ & $m_{23.2}$ & $DP_{8.0/8.4}$ & $DP_{8.4/14.8}$ & $DP_{14.8/23.2}$   \\
\hline
B0023--263&            & G  & 0.3216 & D  &  0.4      &   0.27     &     0.67     &    1.5      &      1.48        &    0.40          &     0.45              \\
B0114--211&            & G  & 1.4100 & Pe &           &            &             &    5.0      &                  &                  &                        \\
B0116+319& 4C31.04    & G  &  0.0602 & P  &  0.27     &   0.18     &     --      &     --      &    1.50          &                  &                        \\
B0127+233& 3C43       & Q  &  1.4590 & T  &   5.3     &    5.5     &     5.2     &    4.9      &    0.96          &   1.06           &     1.06                 \\
B0221+276& 3C67       & G  &  0.3102 & T  &  3.3      &   3.2      &      3.1    &             &    1.03          &    1.02          &                         \\
B0223+341& 4C34.07    & Q  &  2.9100 & D  &   0.43    &  0.34      &     0.58    &     0.85    &    1.26          &     0.59         &   0.68                  \\
B0319+121&            & Q  &  2.6620 & P  & 4.4       &   4.2      &      4.7    &             &    1.05          &     0.89         &                         \\
B0404+767& 4C76.03    & G  &  0.5985 &  P  & 0.2       &    0.06    &     3.7     &             &     3.3          &    0.016         &                        \\
B0531+194&            & G  &         & D  & --         &    --      &      --     &             &     --           &    --            &                          \\
B0538+498& 3C147      & Q  &  0.5450 & D  & 0.93       &   0.98     &     2.9     &             &      0.95        &    0.34          &                         \\
B0941--080&            & G  & 0.2280 & P  &   --      &      --    &    --       &             &      --          &    --           &                         \\
B1005+077& 3C237      & G  &  0.8770 & D  &  0.63       & 0.73       &    3.6      &             &     0.86         &    0.2          &                         \\
B1151--348&            & Q  & 0.2580 & P  &   1.3     &     1.26   &     2.1     &             &     1.03         &     0.6         &                         \\
B1153+317& 4C31.38    & Q  & 0.4170  & D  &   2.1     &      2.1   &             &             &     1.0          &                 &                         \\
B1210+134& 4C13.46    & Q  &  1.1389 &  C  &   6.6     &   6.8      &    6.3      &             &     0.97          &    1.07         &                         \\
B1245--197&            & Q &  1.2750 & Pe &   0.07    &   0.15     &   0.57      &             &    0.47          &    0.26          &                          \\
B1311+678& 4C67.22    & G &          & P  &    --     &      --    &     --      &             &    --            &     --          &                          \\
B1323+321& 4C32.44    & G &  0.3680 & P  &  0.78     &     0.93   &    1.1      &             &    0.84          &    0.85         &                         \\
B1328+254& 3C287      & Q &  1.0550 & P  & 4.2        &     3.9   &    4.2      &             &    1.08          &    0.93         &                        \\
B1345+125& 4C12.50    & G &  0.1217 &  P  &  0.04       &     0.15   &   0.58      &             &    0.27          &    0.26         &                         \\
B1358+624& 4C62.22    & G &  0.4310 &  P  & 0.38       &     0.24   &    0.39     &     --      &     1.58         &     0.62        &                         \\
B1416+067& 3C298      & Q &  1.4373 & T  &  2.6       &       2.5  &    3.0      &    3.2      &     1.04         &   0.83          &      0.94                \\
B1458+718& 3C309.1    & Q &  0.9050 &  P  &    5.3    &    5.3     &     5.3     &    5.6      &     1.0          &     1.0         &    0.95                 \\
B1524--136&            & Q & 1.6870 &  D  &           &            &             &    1.7      &                  &                 &                         \\
B1634+628& 3C343      & G &  0.9880 & P  &  1.2      &   1.2      &    1.3      &             &    1.0           &    0.92         &                         \\
B1638+124& 4C12.60    & G &  1.1520 & P  &   0.21    &  0.21      &    0.63     &    1.7      &    1.0           &   0.33          &     0.37                \\
B1641+173& 3C346      & G &  0.1620 & T  &  4.4      &   4.0      &     3.8     &    3.6      &    1.1           &    1.05         &     1.06                 \\
B1829+290&4C29.56     & G &  0.8420 & T  &  0.6        &     0.50   &    0.8      &     --      &    1.20          &    0.62         &                         \\
B2247+140& 4C14.82    & Q &  0.2346 & P  &  2.9      &  3.0       &    4.5      &    4.5      &    0.97          &      0.67       &      1.0               \\
\hline
\end{tabular}
\normalfont
\smallskip\noindent
\flushleft{\normalsize {
The table is organised as follows:
Column 1: source name; 
2: other name;
3: Optical Identification;
4: Redshift;
5: Radio structure: P point-like, Pe point-like plus estension, D double, 
  T triple, C complex;
6: Polarisation percentage at 8085\,MHz;
7: Polarisation percentage at 8485\,MHz;
8: Polarisation percentage at 14885\,MHz;
9: Polarisation percentage at 23285\,MHz;
10: Depolarisation index, DP,  between 8485 and 8085\,MHz;
11: Depolarisation index,  DP,  between 14885 and 8485\,MHz;
12: Depolarisation index,  DP,  between 23285 and 14885\,MHz.
}}
\end{table*}
In Tables 3--6, the polarisation
parameters derived for the distinct components of each source 
at each observing frequency are reported.  When more than one PA is 
listed in the Tables these refer to regions with different EVPAs.
Further information may be obtained from the on-line images 
(http://db.ira.inaf.it/aj206-fm/).
\tabcolsep0.1cm
\begin{table*}[h]
\tiny
\centering
\caption{Polarisation parameters for components of each source at 8085\,MHz}
\label{tab:par-8085}
\begin{tabular}{llccrlllcllr}
\hline
Name    & Other name & ID & comp & S      & $3\sigma$& S$p$ &3$\sigma$$p$ & $\chi$   & maj    & min    &  PA   \\
        &            &    &      & mJy    & mJy/b  & mJy  & mJy/b     & deg    & arcsec & arcsec & deg    \\
\hline
B0023--263&            & G  & a    & 1329.2 & 0.29      &  5.7 &  0.15       &    53   &  0.16 &  0.05  & 158    \\
        &            &    & b    &  848.2 &           &  3.3 &             &    49   & 0.11 & 0.06 & 129     \\
B0116+319& 4C31.04    & G  &      & 1097.8 & 0.22      &  3.0 &  0.11       &    --4   & 0.10 & 0.02 & 116     \\
B0127+233& 3C43       & Q  & a$_{12}$  &  637.2 & 0.21      &12.8/16.0 &  0.12   &   80/13 & 0.21 & 0.11 & 148    \\
        &            &    & b    &  114.6 &           & 10.7     &         &   --33   &  1.1 & 0.2 & 59       \\
        &            &    & c    &   24.6 &           &  1.4     &         &    44   & 0.33 & 0.26 & 139      \\
B0221+276& 3C67       & G  & a    & 228.8  & 0.08      & 19.6     &  0.09   &   50--80 & 0.58 & 0.19 & 173       \\
        &            &    & b    & 357.4   &           &          &         &         & 0.17 & 0.13 & 176      \\
B0223+341& 4C34.07    & Q  & a    & 1254.0 & 0.18      &  4.6     &  0.12   &    --6   & 0.03 & 0.02 &  43       \\
        &            &    & b    &   66.2 &           &  1.1     &         &   --54   & 0.22 & 0.07  & 81        \\
B0319+121&            &    &      & 1228.1 & 0.15      & 54.0     &  0.10   &    67   &  0.03 & 0.003 & 158       \\
B0404+767& 4C76.03    & E  &      & 2194.9 & 0.27      &  4.4     &  0.14   &   --41   &  0.076 & 0.013 & 43          \\
B0531+194&            & G  & a    & 703.7  & 0.15      &          &  0.09   &         &  0.19 & 0.09 &  139         \\
        &            &    & b    & 751.3  &           &          &         &         &  0.23 & 0.09 &  123       \\
B0538+498& 3C147      & Q  & a    & 4474.6 & 0.36      & 32.2     &  0.10   &   --23   &  0.33 & 0.26 & 57       \\
        &            &    & b    &  393.6 &           & 12.5     &         &   --70   &  0.32 & 0.27 & 24          \\
B0941--080&            & G  &      & 706.0  & 0.12      &          &  0.09   &         &  0.046 & 0.01 & 144          \\
B1005+077& 3C237      & G  & a    & 684.7  & 0.12      &  6.5     &  0.09   &    28   &  0.16 & 0.06 & 78         \\
        &            &    & b    & 462.6  &           &  0.7     &         &   --52   &  0.15 & 0.06 & 87          \\
B1151--348&            & Q  &      & 1749.8 & 0.21      & 22.7     &  0.31   &    20   & 0.11  & 0.05 &  60         \\
B1153+317& 4C31.38    & Q  & a    &  400.0 & 0.08      &  3.4     &  0.09   &   --62   & 0.11 & 0.06 &  8       \\
        &            &    & b    &  214.9 &           &  9.7     &         &    22   & 0.11 & 0.04 &  48      \\
B1210+134& 4C13.46    & Q  & a    &  307.5  & 0.08      & 22.0     &  0.09   &    45   &  0.15 & 0.09 & 15       \\
        &            &    & b    &  200.0 &           &  8.4     &         &    33   &  0.21 & 0.13 & 135       \\
        &            &    & c    &    1.8 &           &          &         &         & 0.68 & 0.19  & 125      \\
        &            &    & d    &    8.8 &           &  1.8     &         &    55   & 0.28 & 0.22  & 165        \\
        &            &    & e    &    6.5 &           &  0.2     &         &   --42   & 0.65 & 0.35  &  1       \\
        &            &    & f    &   15.3 &           &  3.3     &         &   --84/--57 & 0.48 & 0.21 & 134          \\
B1245--197&            & Q &       & 1553.2  & 0.21      & 0.07     &  0.14   &    58     & 0.09 & 0.02 &  87         \\
B1311+678& 4C67.22    & E &       &  608.1 & 0.06      &          &  0.09   &           & 0.06 & 0.02 &  112     \\
B1323+321& 4C32.44    & G &       & 1620.9 & 0.20      & 12.6     &  0.09   &    7      & 0.06 & 0.01 &  134    \\
B1328+254& 3C287      & Q &       & 2239.6 & 0.30      & 92.9     &  0.10   &    --10    & 0.06 & 0.04 & 25      \\
B1345+125& 4C12.50    & G &       & 2291.7 & 0.40      &  1.0     &  0.10   &    42     & 0.06  & 0.015  & 170      \\
B1358+624& 4C62.22    & G  &      & 1182.9 & 0.05      &  4.5 &  0.18        &     25 &  0.03  &  0.01   &  124   \\
B1416+067& 3C298      & Q  & a+b  & 304.1  & 0.08      & 12.1/11.0 &  0.12   &    --55/--7 & 0.25 & 0.12 & 85    \\
        &            &    & c    & 215.2 &           &           &         &            & 0.03 & 0.0  & 46     \\
        &            &    & d    & 388.7 &           &           &         &            & 0.69 & 0.13 & 102    \\
B1458+718& 3C309.1    & Q  & a/a$_1$ & 359.9 & 0.33      & 15.4/11.8  & 0.15   &   --70/--16  & 0.41 & 0.18 & 116      \\
        &            &    & b    &1598.0 &           & 80.2       &        &    49      & 0.087 & 0.044 & 76     \\
        &            &    & c    &  26.4 &           &  0.6       &        &    78      & 0.45 & 0.27 & 67      \\
        &            &    & d    &  83.3 &           &  1.5       &        &   --48      & 0.36 & 0.29 & 99      \\
        &            &    & e    &  32.1 &           &  0.8       &        &   --44      & 0.63 & 0.39 & 162     \\
B1634+628& 3C343      & G  & a    & 825.3  &  0.14     &  9.9       & 0.10   &    80     & 0.09 & 0.06 &  94       \\
        &            &    & b    &   5.4  &           &           &         &           & 0.79 & 0.28 & 59      \\
B1638+124& 4C12.60    & E  &      & 788.1 & 0.13      &  1.72     &  0.09   &    --77     & 0.14 & 0.012 &  130  \\
B1641+173& 3C346      & G  & a$_{123}$ & 331.4 & 0.10      &0.7/17.3/7.6& 0.09   &   0/--36/1  &  --  &  --   &  --    \\
        &            &    & b    & 205.3  &           &            &        &        & 0.015 & 0.0 &  1        \\
        &            &    & c    &  40.7  &           &            &        &        & 3.11 & 0.9 & 54    \\
B1829+290&4C29.56     & G  & a    &  11.4 & 0.10      &            & 0.10   &           &  0.62 & 0.22 &  88      \\
        &            &    & c+b  & 685.1 &           &  4.2       &        &   --28     & 0.077 & 0.014 & 86     \\
        &            &    & d    &   2.2 &           &            &        &           & 0.40 & 0.04 & 62       \\
        &            &    & e    &  10.3 &           &  0.3       &        &   --45     & 0.61 & 0.27 &  56      \\
B2247+140& 4C14.82    & Q  & a+b  & 833.1 & 0.21      &  17.1/7.1  & 0.12   &   --78/15  & 0.21 & 0.07   &  31       \\
\hline
\end{tabular}
\normalfont
\smallskip\noindent
\flushleft{\normalsize {
The table is organised as follows:
Column 1: source name; 
2: other name;
3:  Optical Identification;
4:  component;
5:  integrated component flux density;
6:  three $\sigma$ noise in total-intensity image;
7:  polarised flux density;
8:  three $\sigma$ noise in the polarised-intensity density;
9:  EVPA;
10-12:  deconvolved sizes of the components; major axis, minor axis, and position angle of the
      major axis, respectively.
}}
\end{table*}
\tabcolsep0.1cm
\begin{table*}[h]
\tiny
\centering
\caption{ Polarisation parameters for components of each source at 8485\,MHz}
\label{tab:par-8485}
\begin{tabular}{llccrlllcllr}
\hline
Name    & Other name & ID & comp & S      & $3\sigma$& S$p$ &3$\sigma$$p$ & $\chi$   & maj    & min    &  PA   \\
        &            &    &      & mJy    & mJy/b  & mJy  & mJy/b     & deg    & arcsec & arcsec & deg    \\
\hline
B0023--263&            & G  & a    & 1261.8 & 0.30      &  3.8 &  0.15       &    63   &  0.16 &  0.05  & 157    \\
        &            &    & b    &  893.7 &           &  2.0 &             &    59   & 0.12 & 0.06 & 133     \\
B0116+319& 4C31.04    & G  &      & 1070.8 & 0.30      &  1.97 &  0.10       &    --4   & 0.09 & 0.019 & 116     \\
B0127+233& 3C43       & Q  & a12  &  619.2 & 0.11      & 6.4/20.7 &  0.11   &   80/13 & 0.21 & 0.11 & 147    \\
        &            &    & b    &   98.4 &           & 11.9     &         &   --34   &  1.0 & 0.18 & 58       \\
        &            &    & c    &   25.5 &           &  1.9     &         &    44   & 0.43 & 0.29 & 136      \\
B0221+276& 3C67       & G  & a    & 219.0  & 0.07      & 17.8     &  0.08   &   50--80 & 0.58 & 0.18 & 173       \\
        &            &    & b    & 339.6   &           &          &         &         & 0.16 & 0.13 & 175      \\
B0223+341& 4C34.07    & Q  & a    & 1225.6 & 0.11      &  3.6     &  0.09   &    --4   & 0.03 & 0.014 &  54       \\
        &            &    & b    &   55.2 &           &  0.8     &         &   --55   & 0.019 & 0.085  & 77        \\
B0319+121&            &    &      & 1211.8 & 0.15      & 50.9     &  0.12   &    67   &  0.03 & 0.002 & 160       \\
B0404+767& 4C76.03    & E  &      & 2124.2 & 0.16      &  1.26    &  0.13   &   --60   &  0.076 & 0.007 & 43          \\
B0531+194&            & G  & a    & 624.3  & 0.15      &  0.8     &  0.08   &   --45   &  0.23 & 0.09 &  136         \\
        &            &    & b    & 719.3  &           &  0.4     &         &   --45  &  0.21 & 0.09 &  122       \\
B0538+498& 3C147      & Q  & a    & 4258.0 & 0.39      & 31.6     &  0.15   &   --18   &  0.22 & 0.06 & 55       \\
        &            &    & b    &  370.3 &           & 13.6     &         &   --75   &  0.40 & 0.10 & 18          \\
B0941--080&            & G  &      & 673.5  & 0.12      &          &  0.11   &         &  0.06 & 0.009 & 142          \\
B1005+077& 3C237      & G  & a    & 659.9  & 0.12      &  6.9     &  0.09   &    34   &  0.16 & 0.06 & 78         \\
        &            &    & b    & 439.0  &           &  1.1     &         &   --52   &  0.16 & 0.07 & 88          \\
B1151--348&            & Q  &      & 1678.7 & 0.81      & 21.0     &  0.39   &    26   & 0.11  & 0.04 &  60         \\
B1153+317& 4C31.38    & Q  & a    &  383.1 & 0.06      &  2.6     &  0.08   &   --66   & 0.10 & 0.06 &  5       \\
        &            &    & b    &  203.3 &           &  9.6     &         &    22   & 0.10 & 0.04 &  48      \\
B1210+134& 4C13.46    & Q  & a    &  297.4  & 0.07      & 20.7     &  0.11   &    49   &  0.15 & 0.10 & 11       \\
        &            &    & b    &  177.4  &           &  8.1     &         &    34   &  0.20 & 0.12 & 112       \\
        &            &    & c    &  10.2  &           &          &         &         &  --    &  --     &  --       \\
        &            &    & d    &   10.1 &           &  2.1     &         &    62   & 0.33 & 0.27  & 2        \\
        &            &    & e    &    3.9 &           &  0.3     &         &   --46   & 0.45 & 0.35  &  176       \\
        &            &    & f    &   14.7 &           &  3.7     &         &   --84/--57 & 0.57 & 0.24 & 143          \\
B1245--197&          & Q &       & 1486.7  & 0.17      &  2.2     &  0.13   &    --76     & 0.088 & 0.016 &  87         \\
B1311+678& 4C67.22    & E &       &  583.1 & 0.10      &          &  0.08   &           & 0.06 & 0.02 &  113     \\
B1323+321& 4C32.44    & G &       & 1566.8 & 0.21      & 14.6     &  0.11   &    15      & 0.06 & 0.013 &  133    \\
B1328+254& 3C287      & Q &       & 2167.2 & 0.33      & 85.4     &  0.19   &    --9    & 0.064 & 0.039 & 23      \\
B1345+125& 4C12.50    & G &       & 2215.9 & 0.51      &  3.3     &  0.11   &    --81     & 0.063  & 0.013  & 170      \\
B1358+624& 4C62.22    & G  &      & 1136.5 & 0.14      &  2.7     &  0.12   &     25 &  0.036  &  0.007   &  119   \\
B1416+067& 3C298      & Q  & a+b  & 286.0  & 0.09      & 11.3/10.4 &  0.07  &    --57/--6 & 0.25 & 0.12 & 85    \\
        &            &    & c    & 210.4 &           &           &         &            & 0.04 & 0.0  & 47     \\
        &            &    & d    & 357.2 &           &           &         &            & 0.68 & 0.14 & 103    \\
B1458+718& 3C309.1    & Q  & a/a1 & 348.0 & 0.28      & 14.1/11.4  & 0.15   &   --64/--18  & 0.42 & 0.18 & 116      \\
        &            &    & b    &1543.1 &           & 78.0       &        &    49      & 0.080 & 0.044 & 78     \\
        &            &    & c    &  28.7 &           &            &        &    78      & 0.53 & 0.33 & 172      \\
        &            &    & d    &  75.9 &           &  1.6       &        &   --48      & 0.36 & 0.24 & 96      \\
        &            &    & e    &  41.4 &           &            &        &   --44      & 1.00 & 0.54 & 28     \\
B1634+628& 3C343      & G  & a    & 778.1 &  0.16     &  9.1       & 0.09   &    77     & 0.083 & 0.062 &  100       \\
        &            &    & b    &   3.3 &           &           &         &           & 0.62 & 0.33 & 123      \\
B1638+124& 4C12.60    & E  &      & 757.0 & 0.10      &  1.60     &  0.08   &    --77     & 0.14 & 0.02 &  130  \\
B1641+173& 3C346      & G  & a123 & 311.6 & 0.06      &0.0/15.5/7.6& 0.07   &   --/--36/0  &  --    & --   &  --      \\
        &            &    & b    & 206.7 &           &            &        &        & 0.007 & 0.00 &  37        \\
        &            &    & c    &  43.6 &           &            &        &        & 1.41 & 0.82 & 95    \\
B1829+290&4C29.56     & G  & a    &  9.2 & 0.07      &            & 0.09   &           &  0.45 & 0.22 &  87      \\
        &            &    & c+b  & 652.5 &           &  3.2       &        &   --30     & 0.07 & 0.010 & 87     \\
        &            &    & d    &   1.6 &           &            &        &           & 0.28 & 0.09 & 49       \\
        &            &    & e    &  11.0 &           &  0.2       &        &   --45     & 0.58 & 0.52 &  28      \\
B2247+140& 4C14.82    & Q  & a+b  & 794.9 & 0.36      &  17.2/6.8  & 0.08   &   --79/15  & 0.21 & 0.05   &  31       \\
\hline
\end{tabular}
\normalfont
\smallskip\noindent
\flushleft{\normalsize {
The table is organised as follows:
Column 1: source name; 
2: other name;
3: Optical Identification;
4: component;
5: integrated flux density of the components;
6: three $\sigma$ error on the total-intensity image;
7: polarised flux density;
8: three $\sigma$ error on the polarised-intensity image;
9: EVPA;
10-12: deconvolved sizes of the components; major axis, minor axis, and position angle of the
      major axis, respectively.
}}
\end{table*}
\tabcolsep0.1cm
\begin{table*}[h]
\tiny
\centering
\caption{ Polarisation parameters for components of each source at 14885\,MHz}
\label{tab:par-14885}
\begin{tabular}{llccrlllcllr}
\hline
Name    & Other name & ID & comp & S      & $3\sigma$& S$p$ &3$\sigma$$p$ & $\chi$   & maj    & min    &  PA   \\ 
        &            &    &      & mJy    & mJy/b  & mJy  & mJy/b     & deg    & arcsec & arcsec & deg    \\ 
\hline
B0023--263&            & G  & a12  &  747.70 & 0.54      & 2.9/2.1 &  0.45    &    45/--14 &  0.11 &  0.03  & 150    \\ 
        &            &    & b    &  509.30 &           &  3.4    &          &   --23    & 0.089 & 0.038 & 129     \\ 
B0116+319& 4C31.04    & G  &      &  746.30 & 0.33      &      &  0.38       &         & 0.098 & 0.025 & 114     \\ 
B0127+233& 3C43       & Q  & a1   &  241.80 & 0.30      & 11.4 &  0.34       &    14   & 0.13 & 0.05 & 86          \\ 
        &            &    & a2  &  162.00  &           & 11.7 &             &    81   & 0.12 & 0.0 & 2     \\
        &            &    & b    &   27.90 &           &      &             &         & 0.53 & 0.16 & 55       \\ 
        &            &    & c    &    8.48 &           &      &             &         & 0.28 & 0.07 &  37      \\ 
B0221+276& 3C67       & G  & a12 & 100.7    & 0.21      &  5.4 &  0.23       &    25     & 0.61 & 0.164 & 172       \\
        &            &    & a3   & 8.6    &           &       &             &         &  -- &  --  & --     \\
        &            &    & b   & 178.0    &           &  3.6 &             &   --49   & 0.14 & 0.12 & 168      \\
        &            &    & c   & 1.24     &           &      &             &         & 0.06 & 0.026 & 168      \\
B0223+341& 4C34.07    & Q  & a    &  845.40 & 0.21      &  3.2     &  0.30   &    --70   & 0.017 & 0.006 &  108       \\ 
        &            &    & b    &   25.7 &           &  1.9     &          &   --75   & 0.158 & 0.055  & 88        \\
        &            &    & b0    &  3.38 &           &          &          &         & 0.25 & 0.04  & 59        \\ 
B0319+121&            &    &      &  934.80 & 0.27      & 43.6     &  0.28   &    67   &  0.018 & 0.005 & 164       \\ 
B0404+767& 4C76.03    & E  &      & 1496.00 & 0.27      & 54.7     &  0.32   &    25   &  0.06 & 0.010 & 46          \\ 
B0531+194&            & G  & a    & 294.20  & 0.33      &          &  0.30   &         &  0.124 & 0.084 &  138         \\ 
        &            &    & b    & 365.90  &           &          &         &         &  0.15 & 0.077 &  123       \\ 
B0538+498& 3C147      & Q  & a1    & 1409.70 & 0.42      & 58.1    &  0.42   &    50   &  0.06 & 0.0 & 105       \\ 
        &            &    & a2    & 1029.40 &           &  2.9    &         &    45   &  0.12 & 0.06 & 63       \\
        &            &    & b    &  178.10 &           & 15.1     &         &    90   &  0.21 & 0.08 & 16          \\ 
B0941--080&            & G  &      & 390.40  & 0.24      &          &         &         &  0.065 & 0.008 & 143          \\ 
B1005+077& 3C237      & G  & a    & 336.20  & 0.18      &  15.9     &  0.30   &  2/42   &  0.16 & 0.04 & 78         \\ 
        &            &    & b    & 215.00  &           &  4.0      &         &   0    &  0.12 & 0.05 & 85          \\ 
        &            &    & b1   &   3.86  &           &           &         &        &  0.14 & 0.04 & 105          \\
B1151--348&            & Q  &      &  966.40 & 0.68      & 20.0     &  0.64   &     9   & 0.10  & 0.04 &  68         \\ 
B1210+134& 4C13.46    & Q  & a    &  205.8  & 0.18      & 20.2     &  0.30   &     40  &  0.08 & 0.05 & 42       \\ 
        &            &    & b    &  116.3  &           &  2.3     &         &    35   &  0.04 & 0.03 & 21       \\ 
        &            &    & c    &   26.4? &           &          &         &         &  --   & --   & --        \\ 
        &            &    & d    &   3.20  &           &          &         &         & 0.49 & 0.23  & 137      \\ 
        &            &    & f    &    5.30 &           &          &         &           & 0.40 & 0.15 & 141          \\ 
B1245--197&            & Q & a     &  804.9  & 0.24      &  5.0     &  0.32  &     52     & 0.0 & 0.0 &           \\ 
        &            &   & b     &   71.8  &           &          &         &            & 0.11 & 0.02 &  102         \\ 
B1311+678& 4C67.22    & E &       &  328.30 & 0.15      &          &  0.32   &           & 0.05 & 0.02 &  113     \\ 
B1323+321& 4C32.44    & G &       & 1038.00 & 0.27      & 11.4     &  0.19   &    11      & 0.06 & 0.010 &  132    \\ 
B1328+254& 3C287      & Q &       & 1389.0 & 0.42       & 58.8     &  0.25   &    --4    & 0.05 & 0.03 & 28      \\ 
B1345+125& 4C12.50    & G &       & 1535.0 & 0.18      &  8.9      &  0.25   &     20   & 0.06  & 0.01  & 171      \\ 
B1358+624& 4C62.22    & G  &      &  721.04 & 0.45      &  2.8     &  0.32   &     45 &  0.029  &  0.011   &  96   \\ 
B1416+067& 3C298      & Q  & a    &  85.40  & 0.25      &  6.1     &  0.32  &    --63     & 0.12 & 0.07 & 33    \\ 
        &            &    & b    &  46.3   &           &  6.1     &        &     --2     & 0.19 & 0.07 & 71      \\ 
        &            &    & c1    &  70.50 &           &  0.7     &         &    --1     & 0.11 & 0.03  & 96     \\ 
        &            &    & c2    & 167.20 &           &          &         &           & 0.014 & 0.0  & 54     \\ 
        &            &    & d    &  60.10 &           &           &         &            & 0.08 & 0.07 & 41     \\ 
B1458+718& 3C309.1    & Q  & a    &  76.60 & 0.36      &    7.7    & 0.43    &   --87/--46  & 0.16 & 0.0 & 13      \\ 
        &            &    & a0   & 104.7  &           &           &         &           & 0.39  & 0.21 & 110     \\ 
        &            &    & a1   &  38.3  &           &  3.3      &         &   --23     & 0.022 & 0.096 & 95    \\ 
        &            &    & b    & 1051.90 &          & 58.2       &        &    45     & 0.048 & 0.025 & 146     \\ 
        &            &    & d    &  28.00  &          &            &        &           & 0.28 & 0.15 & 89      \\ 
B1634+628& 3C343      & G  & a    & 376.00  &  0.18     &  5.1      & 0.32   &    42     & 0.07 & 0.05 &  87   \\
B1638+124& 4C12.60    & E  & a    & 432.10  & 0.23      &   0.7     &  0.30   &   --1    & 0.02 & 0.01 &  122  \\ 
        &            &    & b    &  91.90 &           &   2.6     &         &    90     & 0.04 & 0.02 &  172  \\ 
B1641+173& 3C346      & G  & a12  & 171.30 & 0.15      & 1.2/14.0  & 0.34    &    --35    & 0.14 & 0.08 & 95     \\ 
        &            &    & a3   &  10.20 &           & 1.2       &         &   --1      & 0.18 & 0.09 &  115   \\ 
        &            &    & b    & 239.70  &           &            &        &        & 0.005 & 0.003 & 133        \\ 
B1829+290&4C29.56     & G  & a    &  --     & 0.24      &            & 0.30   &       & --   &  --   &  --        \\
        &            &    & c    & 356.70 &           &  2.9       &        &    52     & 0.06 & 0.01 & 86     \\ 
        &            &    & c1    &  2.60 &           &            &        &           & 0.20 & 0.02 & 51     \\
        &            &    & d    &   --   &           &            &        &           & --   & --    &  --       \\ 
B2247+140& 4C14.82    & Q  & a    & 376.30 & 0.24      &  18.5      & 0.32   &   --85  & 0.07 & 0.03   &  63       \\ 
        &            &    & b    & 207.40 &           &  7.9       &        &    13     & 0.13 & 0.06   &  162     \\
\hline
\end{tabular}
\normalfont
\smallskip\noindent
\flushleft{\normalsize {
The table is organised as follows:
Column 1: source name; 
2: other name;
3: optical identification;
4: component;
5: integrated flux density of the components;
6: three $\sigma$ error on the total-intensity image;
7: polarised flux density;
8: three $\sigma$ error on the polarised-intensity image;
9: EVPA;
10-12: deconvolved sizes of the components; major axis, minor axis, and position angle of the
      major axis, respectively.
}}
\end{table*}
\tabcolsep0.1cm
\begin{table*}[h]
\tiny
\centering
\caption{ Polarisation parameters for the components of each source at 23285\,MHz}
\label{tab:par-23285}
\begin{tabular}{llccrlllcllr}
\hline
name    & other name & ID & comp & S      & $3\sigma$& S$p$ &3$\sigma$$p$ & $\chi$   & maj    & min    &  PA   \\ 
        &            &    &      & mJy    & mJy/b  & mJy  & mJy/b     & deg    & arcsec & arcsec & deg    \\ 
\hline
B0023--263&            & G  & a12  &  489.3 & 0.42      &  3.2   &  0.70    &    43    &  0.07 &  0.01  & 152    \\ 
        &            &    & b    &  316.8 &           &  8.6   &          &     34    & 0.06 & 0.03 & 131     \\
B0114--211&            & G  &      &  141.3  &  0.27     &  7.1   &  0.47    &     1     & 0.054 & 0.019  &  112   \\
B0116+319& 4C31.04    & G  &      &  550.2 & 0.27      &      &  0.30       &         & 0.10 & 0.026 & 112     \\ 
B0127+233& 3C43       & Q  & a1   &  113.0 & 0.24      &  5.0 &  0.28       &     3   & 0.14 & 0.02  & 178     \\ 
        &            &    & a2   &   79.2 &           &  5.6 &             &    48   & 0.08 & 0.05  & 124     \\
        &            &    & b    &   25.7 &           &      &             &         &   -- & --    &  --       \\ 
B0223+341& 4C34.07    & Q  & a    &  610.2 & 0.21      &  5.3     &  0.36   &     1   & 0.05 & 0.005 &  114       \\ 
        &            &    & b    &   12.4 &           &           &         &         & 0.126 & 0.052  & 82        \\ 
        &            &    & b0    &   3.1 &           &           &         &         & 0.09 & 0.09  & 51        \\ 
B1358+624& 4C62.22    & G  &      &  486.0 & 0.30      &           &  0.43   &        &  0.03  &  0.003   &  118   \\  
B1416+067& 3C298      & Q  & a    &  36.8  & 0.24      &  1.0     &  0.36  &    --70     & 0.13 & 0.08 & 65    \\  
        &            &    & b    &  16.1  &           &  0.96    &        &     1      & 0.16 & 0.06 & 81      \\  
        &            &    & c1    & 42.1  &           &  3.0     &         &    23     & 0.16 & 0.03  & 114     \\  
        &            &    & c2    & 83.3  &           &  0.8     &         &    73/37  & 0.04 & 0.008  & 118     \\  
        &            &    & d    &  18.9  &           &  0.4     &         &    --88    & 0.09 & 0.06 & 85     \\  
B1458+718& 3C309.1    & Q  & a    &  92.1 & 0.48      &    5.4     & 0.43    &    90/--25 & 0.19 & 0.16 & 161      \\  
        &            &    & a0   &        &           &           &         &             & --    & --    &   --      \\  
        &            &    & a1   &  26.5   &           & 2.6       &         &   --35    & 0.023 & 0.08    & 93    \\  
        &            &    & b    &  767.6 &          & 41.9       &        &    46     & 0.048 & 0.020 & 151     \\  
        &            &    & d    &  11.8  &          &            &        &           & 0.23 & 0.12 & 92      \\  
B1524--136&            & Q  & a    & 279.0  & 0.30     &  4.9      &  0.47   &   65     & 0.10 & 0.012 & 163    \\
        &            &    & b    &  35.3  &          &   0.6      &         &   88     & 0.06 & 0.05 &  74    \\
B1638+124& 4C12.60    & E  & a    & 304.4  & 0.21      &   3.3     &  0.36   &    81    & 0.0 & 0.0 &  0  \\  
        &            &    & b    &  56.5 &           &   2.8     &         &    76     & 0.06 & 0.02 &  142  \\  
B1641+173& 3C346      & G  & a12  &  79.8 & 0.30      & 1.3/10.4  & 0.38    &    -43    & 0.12 & 0.07 & 85     \\  
        &            &    & a3   &   5.1 &           & 1.3        &         &   -1      & 0.12 & 0.08 &  163   \\  
        &            &    & b    & 234.2  &         &            &         &           & 0.003 & 0.0  & 106        \\  
        &            &    & c    &  43.6  &           &            &        &        & 1.41 & 0.82 & 95    \\ 
B1829+290&4C29.56     & G  & c    & 182.1 &           &            &        &           & 0.059 & 0.008 & 87     \\  
B2247+140& 4C14.82    & Q  & a    & 257.5  & 0.18      &  14.8      & 0.32   &   60/--76  & 0.074 & 0.024   &  56       \\  
        &            &    & b    & 159.5  &           &  3.8       &        &    14     & 0.21 & 0.06   &  17     \\ 
\hline
\end{tabular}
\normalfont
\smallskip\noindent
\flushleft{\normalsize {
The table is organised as follows:
Column 1: source name; 
2: other name;
3: optical identification;
4: component;
5: flux density;
6: three $\sigma$ error on the total-intensity image;
7: polarised flux density;
8: three $\sigma$ error on the polarised-intensity image;
9: EVPA;
10-12: deconvolved sizes of the components; major axis, minor axis, and position 
angle of the major axis, respectively.
}}
\end{table*}

Fractional polarisations, depolarisation indices, and Rotation Measure values
for the individual components of each source are reported in
Table~\ref{tab:par-RM}. The Rotation Measure is defined as 
RM$=\Delta\chi / \Delta(\lambda^2)$ in $rad\,m^{-2}$
where $\chi$ is the PA at wavelength $\lambda$, allowing for $n\pi$ 
with $n$ an integer,
in the individual PAs to get the best fit.
When three frequencies are available, the ambiguity implied by the integer 
$n$ can be resolved. The RMs are estimated by fitting the points 
with a linear least-squares fit. The Rotation Measure in the source rest frame 
is defined as RM$_{rf}$=RM$\times(1+z)^2$.
 
We note that polarimetric parameters are derived from images at the full
resolution achieved at the various frequencies. The $\chi$
values are those associated with the polarised emission peaks. The errors
associated with the $\chi$ values used in the $\chi - \lambda^2$ plots
are calculated considering the dispersion of $\chi$ values in boxes of
$5\times5$ pixels around the positions of maximum polarised emission. 

RM plots
are also available at the web site http://db.ira.inaf.it/aj206-fm/. 
Examples of RM plots are presented in Fig.~\ref{fig:rotmplots}. 
\vspace{-0.3cm}
\begin{figure*}[t]
\addtocounter{figure}{+0}
\centering
\includegraphics[width=6cm,angle=-90]{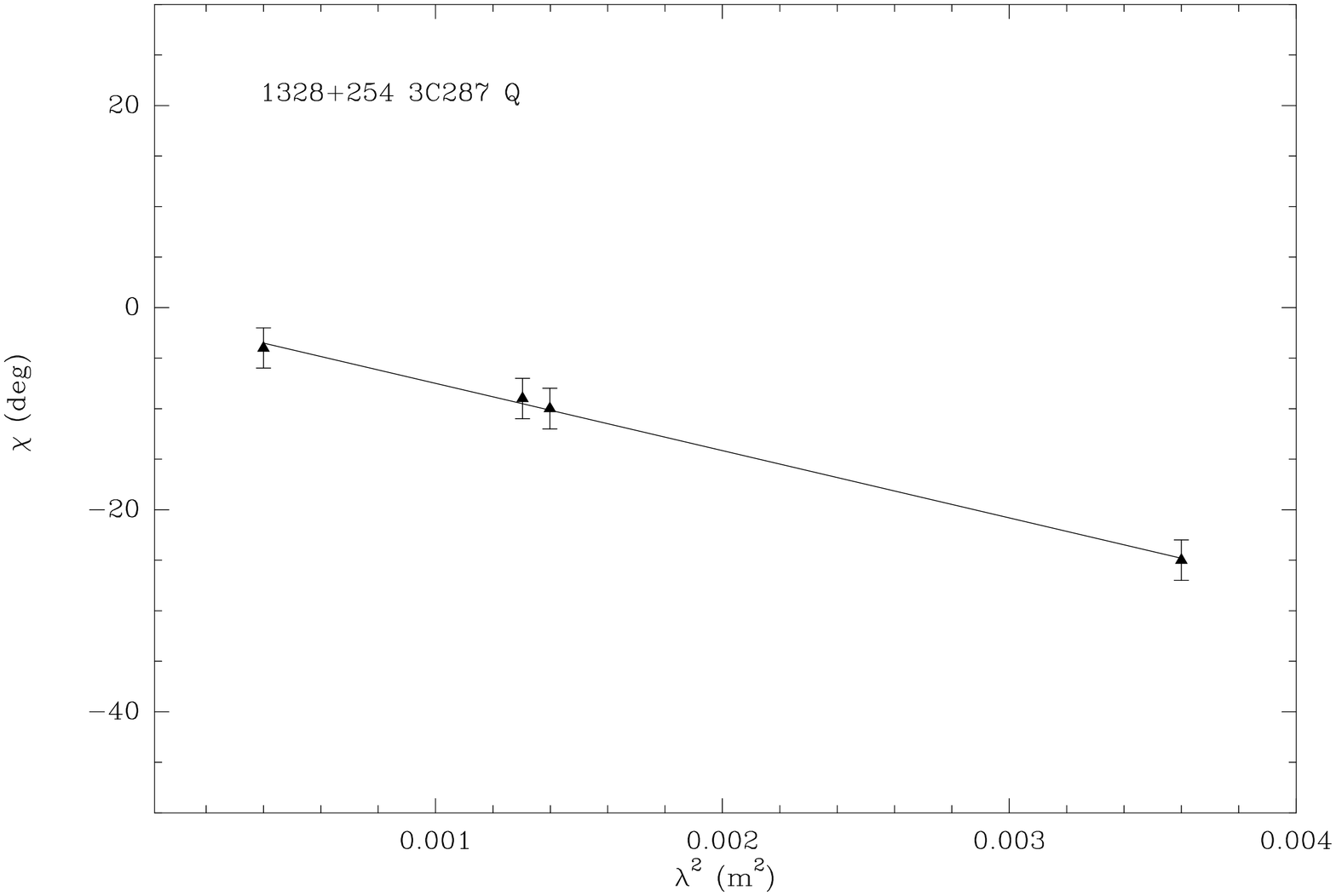}
\includegraphics[width=6cm,angle=-90]{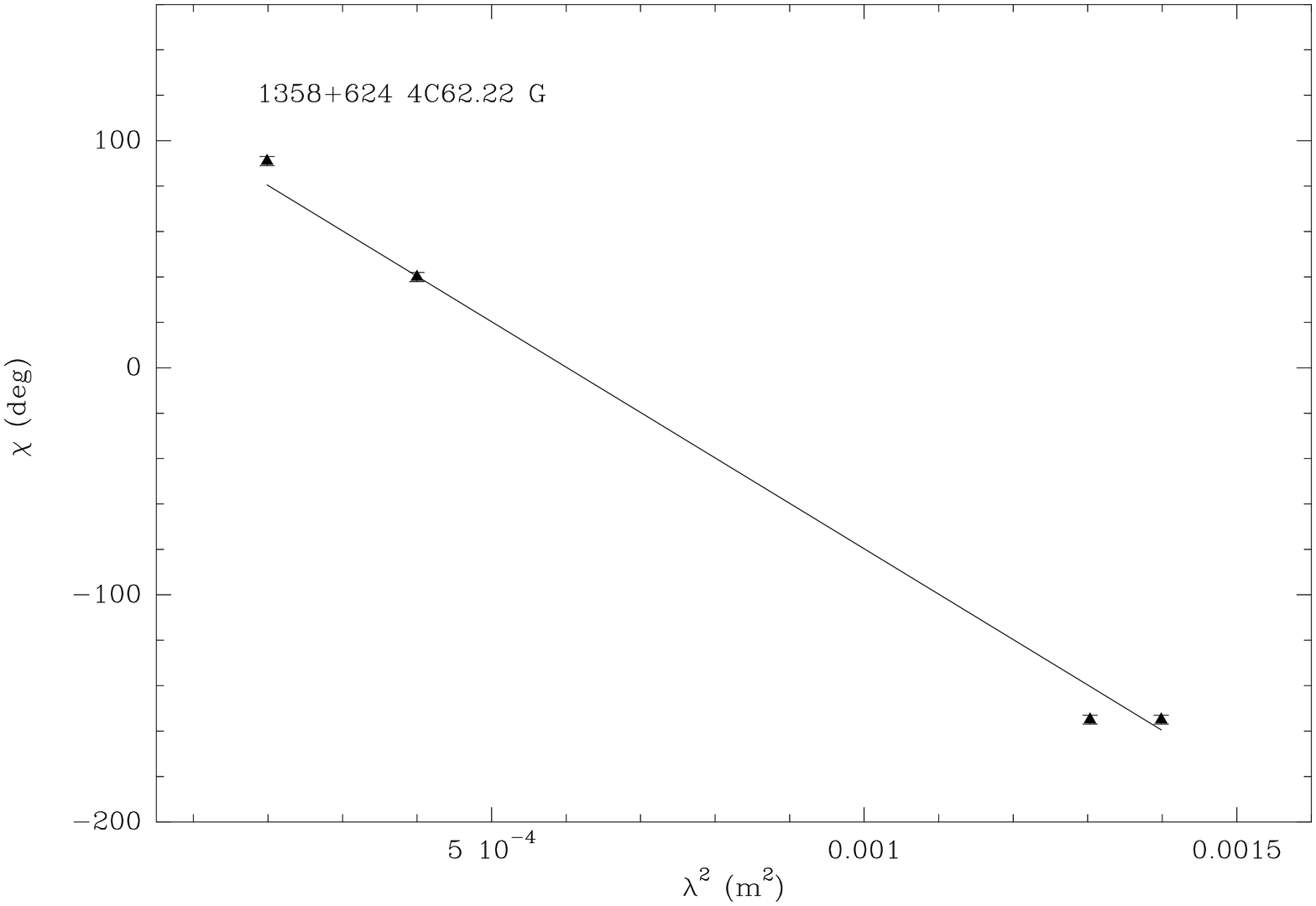}
\caption{Plots of the observed  $\chi$ values of the quasar 1328+254 (3C287)
and of the galaxy 1358+624 (4C62.22) as a function of  $\lambda^2$ for the available 
frequencies.
\label{fig:rotmplots}}
\end{figure*}
\tabcolsep0.1cm
\begin{table*}[h]
\tiny
\centering
\caption{Fractional polarisation, depolarisation index and Rotation Measure
for individual components of each source.}
\label{tab:par-RM}
\begin{tabular}{llccrlllcllrr}
\hline
name    & other name & ID & comp & $m_{8.0}$ & $m_{8.4}$ & $m_{14.8}$ & $m_{23.2}$ & $DP_{8.0/8.4}$ & $DP_{8.4/14.8}$ & $DP_{14.8/23.2}$ & RM       & RM$_{rf}$  \\
        &            &    &      &   [\%]    &   [\%]    &  [\%]      &    [\%]    &                &                  &                  &rad m$^{-2}$&rad m$^{-2}$    \\
\hline
B0023--263&            & G  & a    & 0.43 & 0.30 & 0.67 &  0.65 & 1.43 & 0.45 & 1.03 &  146    &  255    \\
        &            &    & b    & 0.39 & 0.22 & 0.67 &  2.71  & 1.64 & 0.33 & 0.25 &         &         \\
B0114--211&            & G  &      &      &      &      &  5.02 &      &      &      &  646    & 3752    \\
B0116+319& 4C31.04    & G  &      & 0.27 & 0.18 &      &       & 1.50 &      &      &         &         \\
B0127+233& 3C43       & Q  & a1   & 4.52 & 4.38 & 4.71 &  4.1  & 1.03 & 0.93 & 1.67 &  --24   & --145   \\
        &            &    & a2   &      &      & 7.22 &  7.07 &      &      & 1.02 &   --23  & --139   \\
        &            &    & b    & 9.34  & 12.09 &      &       & 0.72 &      &      &         &         \\
        &            &    & c    & 5.69 &  7.45 &      &       & 0.76 &      &      &         &         \\
B0221+276& 3C67       & G  & a1   & 8.57 &  8.13 & 5.36 &       & 1.05 & 1.51 &      &  768    &  1318   \\
        &            &    & a2   &      &      &      &       &      &      &      &  942    &  1617   \\
        &            &    & b    &      &      & 2.02 &       &      &      &      & 1745    &  2995   \\
B0223+341& 4C34.07    & Q  & a    & 0.37 & 0.29 & 0.38 &  0.87 & 1.28 & 0.76 & 0.44 & --2373   & --36279   \\
        &            &    & b    & 1.66 & 1.45 & 7.39 &       & 1.14 & 0.20 &      &   384    & 5871   \\
B0319+121&            &    &      & 4.40 & 4.20 & 4.66 &       & 1.05 & 0.90 &      &  0   &  0   \\
B0404+767& 4C76.03    & E  &      & 0.20 & 0.06 & 3.66 &       & 3.33 & 0.02 &      &  2090   & 5350   \\
B0531+194&            & G  & a    &      & 0.13 &      &       &      &      &      &         &         \\
        &            &    & b    &      & 0.06 &      &       &      &      &      &         &         \\
B0538+498& 3C147      & Q  & a1   & 0.72 & 0.74 & 4.12 &       & 0.97 & 0.18 &      &         &         \\
        &            &    & a2   &      &      & 0.28 &       &      &      &      &         &         \\
        &            &    & b    & 3.18 & 3.67 & 8.48 &       & 0.87 & 0.43 &      &         &         \\
B1005+077& 3C237      & G  & a    & 0.95 & 1.05 & 4.73 &       & 0.90 & 0.22 &      & --174   & --615    \\
        &            &    & b    & 0.15 & 0.25 & 1.86 &       & 0.60 & 0.13 &      & --942   & --3329     \\
B1151--348&            & Q  &      & 1.3  & 1.25 & 2.1  &       & 1.04 & 0.60 &      &  0    &  0    \\
B1153+317& 4C31.38    & Q  & a    & 0.85 & 0.68 &      &       & 1.25 &      &      &  453    &  913    \\
        &            &    & b    & 4.51 & 4.72 &      &       & 0.96 &      &      &  174    &  351    \\
B1210+134& 4C13.46    & Q  & a    & 7.15 & 7.00  & 9.82&       & 1.02 & 0.71 &      & 140    & 640   \\
        &            &    & b    & 4.20 & 4.60  & 1.98 &       & 0.91 & 2.32 &      & --35   & --160        \\
        &            &    & d    & 20.45& 20.79&      &       & 0.98 &      &      &         &         \\
        &            &    & e    & 3.08 & 7.69 &      &       & 0.40 &      &      &          &        \\
        &            &    & f    & 21.57& 25.17& 17.0 &       & 0.86 & 1.48 &      &   349    &  1597    \\
B1245--197&            & Q  & a    & 0.07 &  0.15 & 0.62 &       & 0.47 & 0.24 &      &          &          \\
B1323+321& 4C32.44    & G &       & 0.78 & 0.93 & 1.10 &       & 0.84 & 0.85 &      & --35   & --66   \\
B1328+254& 3C287      & Q &       & 4.15 & 3.94 & 4.23 &       & 1.05 & 0.93 &      & --131 & --556 \\
B1345+125& 4C12.50    & G &       & 0.04 & 0.15 & 0.58 &       & 0.27 & 0.26 &      &--5585   &--8271 \\
B1358+624& 4C62.22    & G  &      & 0.38 & 0.24 & 0.39 &       & 1.58 & 0.61 &      & --3665   & --7496   \\
B1416+067& 3C298      & Q  & a    & 7.60 & 7.59 & 7.14 &  2.72 & 1.00 & 1.06 & 2.63 & --140   & --833  \\
        &            &    & b    &      &      & 13.17 &  5.96 &      &      & 2.21 & --175   & --1042     \\
        &            &    & c1   &      &      & 0.99 &  7.13 &      &      & 0.14 &         &         \\
        &            &    & c2   &      &      &      & 0.96  &      &      &      &         &         \\
        &            &    & d    &      &      &      & 2.12  &      &      &      &         &      \\
B1458+718& 3C309.1    & Q  & a    & 7.56 & 7.33 & 10.05& 5.86  & 1.03 & 0.73 & 1.71 & --1047  & --3920   \\
        &            &    & a1    &     &      & 8.62 & 9.81  &      &      & 0.88 &   314   &  1146    \\
        &            &    & b    & 5.02 & 5.05 & 5.53 & 5.46  & 0.99 & 0.91 & 1.01 &         &        \\
        &            &    & c    & 2.27  &     &      &       &      &      &      &         &         \\
        &            &    & d    & 1.80 & 2.11 & 1.8  &       & 0.85 &      &      & 2792    & 10185   \\
        &            &    & e    & 2.49 &      &      &       &      &      &      &         &         \\
B1524--136& OQ172     & Q  & a    &      &      &      & 1.76  &      &      &      & 2827   & 20411    \\
         &           &    & b    &      &      &      & 1.70  &      &      &      &        &          \\
B1634+628& 3C343      & G  & a    & 1.20 & 1.17 &  1.36&       & 1.03 & 0.86 &      & 698     & 2764    \\
B1638+124& 4C12.60    & E  & a    & 0.22 & 0.21 & 0.16 & 1.08  & 1.05 & 1.31 & 0.15 & 3141   & 14546  \\
        &            &    & b    &      &      & 2.83 & 4.96  &      &      & 0.57 &         &         \\
B1641+173& 3C346      & G  & a12  & 7.72 & 7.41 & 8.87 & 14.66  & 1.04 & 0.84 & 0.60 &   54    &    73   \\
        &            &    & a3   &      &      & 11.76& 25.49 &      &      & 0.46 &   17    &    23   \\
B1829+290&4C29.56     & G  & c    & 0.61 & 0.49 & 0.81 &       & 1.24 & 0.60 &      &--1466   & --4963  \\
        &            &    & e    & 2.91 & 1.82 &      &       & 1.60 &      &      &         &         \\
B2247+140& 4C14.82    & Q  & a    & 2.90 & 3.02 & 4.92 & 5.75  & 0.96 & 0.61 & 0.85 &   209   &  319    \\
        &            &    & b    &      &      & 3.81 & 2.38   &      &      & 1.60 &   454   & 692     \\
\hline
\end{tabular}
\normalfont
\smallskip\noindent
\flushleft{\normalsize {
The table is organised as follows:
Column 1: source name; 
2: other name;
3: optical identification;
4: component;
5: percentage of polarised emission at 8085\,MHz;
6: percentage of polarised emission at 8485\,MHz;
7: percentage of polarised emission at 14885\,MHz;
8: percentage of polarised emission at 23285\,MHz;
9: depolarisation index between 8485\,MHz and 8085\,MHz;
10: depolarisation index between 14885\,MHz and 8485\,MHz;  
11: depolarisation index between 23285\,MHz and 14885\,MHz;
12: Rotation Measure;
13: Rotation Measure in the source rest frame RM$_{rf}$=RM$\times(1+z)^2$.
}}
\end{table*}
\subsection{Comments on individual sources}

Some of the CSS sources presented here have been included in
other sub-arcsecond resolution observations. Parameters from those
studies, and in particular the $\chi$ values, have been
extracted when appropriate.

\noindent
{\bf 0023$-$263 (OB238) --}
This source have been observed at the S/X bands using VLBI by
\citet{tasso02} with the SHEVE array. We confirm the double source
structure of the source seen at both frequencies.

\noindent
{\bf 0114$-$211, (OC224) --}
This source was observed with the VLA A-array at 5, 8.4, and
15\,GHz by \citet{Mantovani94} who reported a triple structure. 
Polarised emission was detected only for the western component
``c''. The rest-frame RM was calculated in that paper assuming a
conventional value of $z$=0.5. With the present 23.2\,GHz observations
we detect total intensity and polarised emission only from component
``c''. Here we have used a redshift of 1.41$\pm$0.05 
(McCarthy et al. 1996) for estimating the rest-frame RM. 

\noindent
{\bf 0116$+$319 (4C31.04) --}  
This source was observed by \citet{breugel84} at 15\,GHz with the
VLA. They detected no polarised emission above their detection
limit of 1.5\% of the flux density of 1226$\pm$16 mJy, which is a
factor of $\sim$1.6 higher than the flux density we find at the same
frequency.

\noindent
{\bf 0127$+$233 (3C43) --}
The subject of many investigations, 3C43 shows a peculiar structure
characterized by a sharp bend of more than 90$^\circ$ in the jet.
Polarised emission has been reported by \citet{breugel92},
\citet{A-G95}, \citet{Sanghera95}, \citet{Ludke98} and 
Cotton et al. (2003c) at sub-arcsecond
resolution.  The EVPA in all these images made from data taken at
different frequencies, is perpendicular to the jet axis, nicely
following the bend in the jet.  Small values of the RM are found along
the jet region, while a high value of RM is measured for the
northern component.  VLBI observations were reported by
\citet{Nan91b}, \citet{Spencer91}, and \citet{Fanti02}. Polarimetric
VLBI observations were performed by \citet{Mantovani03} at 8.4\,GHz.
These authors detected polarised emission along the jet with
polarisation percentages up to 12\%. However, no polarisation was
detected from the region of the core. A similar result was obtained
at 1.6\,GHz by \citet{Cotton03}. Both articles report a large RM
near the bend in the jet.

\noindent
{\bf 0221$+$276 (3C67) --}
This source shows a triple structure from our 15\,GHz observations, as
did MERLIN observations at 5\,GHz by \citet{Sanghera95}.  Because of
its almost-flat spectrum, we confirm that the central weak component
close to the southern lobe 
is the core of the source, as suggested by \citet{Sanghera95} and 
\citet{Ludke98}. Of the two
lobes, that to the north is strongly polarised at the frequencies we
observed.  The area with the highest polarised emission shows RMs
of up to 1620 rad m$^{-2}$ in the source rest frame.  The southern
lobe is 2\% polarised at 15\,GHz, and  depolarises quickly towards
lower frequencies. Polarised emission from this component is only
marginally detected by our X-band observations. The RM value
($\sim$3000 rad m$^{-2}$ in the source rest-frame) calculated for this
component is rather uncertain.

\noindent
{\bf 0223$+$341 (4C34.07) --}
The images at 15 and 23.2\,GHz reveal a triple structure. The central
component is unpolarised and has an almost flat spectrum between the
two frequencies. In contrast, both lobes are polarised. However,
the less-polarised northern component shows depolarisation between
23.2 and 15\,GHz, but a constant polarisation percentage between 15
and 8.0\,GHz. The source rest-frame RM value found for this
component of more than --36,000 rad m$^{-2}$ is calculated after
adding 180$^\circ$ to $\chi$ at 23.3 \,GHz.
Subtracting $180^\circ$ from $\chi$ at 8.0 and 8.4\,GHz will
generate an even higher RM. Both solutions give acceptable fits
to the $\lambda^2$ law.  If we do not adjust any of the above three
$\chi$ values, a rare case is implied in which the $\chi$ values do
not obey the $\lambda^2$ law.

The southern component is strongly polarised at 15\,GHz, while the
polarisation percentage is a factor of 5 lower at X-band. An RM value
of RM=5871 rad m$^{-2}$ in the source rest-frame can be unambiguosly
derived.

\noindent
{\bf 0319$+$121 --}
This source is point-like at 15\,GHz, but shows an extension to the
north at X-band. VLBI S/X band images from the US Naval
Observatory Radio Reference Frame Image Database 
(http://rorf.usno.navy.mil/RRFID/) show a core-jet
structure, with the jet in the same direction we find. The source
presents little depolarisation from higher to lower frequencies and
an RM of zero.

\noindent
{\bf 0404$+$767 (4C76.03) --}
Again a point-like source at sub-arcsecond resolution. In this case
there is a strong depolarisation between 15 and 8.0\,GHz. However,
the percentage polarisation at 8.4\,GHz is much lower that at 8.0\,GHz,
which is possibly due to uncertainties caused by the low levels of
polarization. Nevertheless, the
$\chi$ values fit the $\lambda^2$ law nicely, with RM=5350 
rad~m$^{-2}$ in the source rest-frame.  The VLBI image at 8.0\,GHz by
\citet{Xu95} shows a triple structure for 0404$+$767 at milli-arcsecond
resolution.

\noindent
{\bf 0531$+$194 --}
A compact double source that is unpolarised at the detection levels of
all our observations.

\noindent
{\bf 0538$+$498 (3C147) --}
Our results for this source are fully described in \citet{Junor99a}.

\noindent
{\bf 0941$-$080 --}
A point-like source with polarised emission below the detection limits
at all of the available frequencies.

\noindent
{\bf 1005$-$077 (3C237) --}
The triple structure seen by \citet{Ludke98} observing 3C237 with
MERLIN at 5\,GHz is confirmed by our 15\,GHz observations.  The
central component has a flat spectral index between 5 and
15\,GHz, has no detectable polarised emission, and is almost
certainly the core of the source. Both lobes are polarised at
X- and U-band, showing strong depolarisation between 15 and
8.0\,GHz.  The lobes are completely depolarised by 5\,GHz as reported
by \citet{Ludke98}.  The eastern and western lobes have RMs of 942
and 615 rad m$^{-2}$ respectively.

3C237 has also been observed by \citet{A-G95} at 8.4\,GHz and by
\citet{breugel92} at 15\,GHz. 

\noindent
{\bf 1151$-$348 --}
This is a point-like source at sub-arcsecond resolution. It is up to
2.1\% polarised at 15\,GHz and shows depolarisation towards
X-band, with an RM of about zero. Dual-band S/X VLBI observations made by
\citet{tasso02} revealed that 1151$-$348 has a double structure with the
components separated by about 90 milli-arcseconds.

\noindent
{\bf 1153$+$317 (4C31.38) --}  
Our observations at 8~GHz show this to be a double source, with both components
being polarised.  From the observations of \citet{Lonsdale93}, it is
possibly a triple source at 15\,GHz. Lonsdale et al. (1993) also observed
the source at 5\,GHz. Adopting their polarisation measurements, we
compute RMs of 913  and 351 rad m$^{-2}$ for components {\bf a} and
{\bf b} respectively.

\noindent
{\bf 1210$+$134 (4C13.46) --}
This source presents a very complex structure, dominated by the
northern component in which two blobs of emission, both strongly
polarised, are embedded in an extended region of weak emission 
(Fig. 1). A three-arcsecond long jet extends south, beginning from the
south-western most of the two blobs mentioned above.  Polarised
emission is detected all along the jet at X-band. The south-western
and north-eastern blobs
show RM=640 and $-160$ rad m$^{-2}$ respectively. The values we
find for the percentage polarisations are rather puzzling:  the
brightest component {\bf a} depolarises going from 15 to 8.4\,GHz
($DP_{8485/14885}=0.7$), while component {\bf b} strongly repolarises
($DP_{8485/14885}=2.3)$). Polarised emission is also detected at the
three observing frequencies for component {\bf f} at the end of the
jet, which also shows repolarisation and has RM=1597 rad m$^{-2}$.

\noindent
{\bf 1245$-$197 --}
This source is slightly resolved at 15\,GHz. The image shows an
extension to the west of the brightest point-like component. This
component is weakly polarised, and has a very large depolarisation
index.  It is one of the few cases that does not follow the
$\lambda^2$ law.  1245$-$197 was also observed with the VLA at
1.36\,GHz by \citet{Stang05} who found a component about 4 arcmin
to the west of the main component.

\noindent
{\bf 1311$+$678, 4C67.22 --}
Polarised emission is not detected for this source which is
slightly extended to the south-east at 15\,GHz.  1311$+$678 shows
double structure at milli-arcsecond resolution \citep [see] [] {Xu95}.

\noindent
{\bf 1323$+$321 (4C32.44) --}
Again a point-like source, it is slightly depolarised between 15\,GHz
and X-band. However, the percentage polarisation is low.  The value
of the intrinsic RM is also low (RM$_{\it rf}$=--66 rad m$^{-2}$).
1323$+$321 is part of the MOJAVE monitoring programme which is
performed with the VLBA at 15\,GHz.  There it shows a double
structure \citep{Lister05} with the two components being separated
by about 50 milli-arcsec.

\noindent
{\bf 1328$+$254 (3C287) --}
This point-like source is about 4\% polarised, with little
depolarisation from higher to lower frequencies. The intrinsic
RM$_{\it rf}$ is --556 rad m$^{-2}$.  3C287 was observed by
\citet{Kellermann98} with the VLBA at 15\,GHz and showed a point-like
structure with a weak extension to the south.

\noindent
{\bf 1345$+$125 (4C12.50) --}
A point-like source that is only weakly polarised. It presents a very
high depolarisation index going from higher to lower frequency and a
very large intrinsic RM (RM$_{\it rf}$=--8271 rad m$^{-2}$). VLBI
observations by \citet{Stang97} at 5\,GHz and \citet{Kellermann98}
at 15\,GHz reveal a complex, long, thin structure for the source.

\noindent
{\bf 1358$+$624, (4C62.22) --}
The present observations show a point-like structure for 1358$+$624.
It is weakly polarised, and presents noticeable depolarisation,
with a large intrinsic RM (RM$_{\it rf}$=--7496 rad m$^{-2}$). VLBI
observations at 18\,cm by \citet{Dallacasa95} show a double structure
with a faint bridge, identified as a jet, connecting the two lobes.

\noindent
{\bf 1416$+$067, (3C298) --}
This source extends east-west, and is resolved into multiple
components. We can identify at least five of these in our
23.2\,GHz image, which shows percentage polarisations ranging
from about 1\% to 7\% at 23.2\,GHz. Polarised emission was detected
only on the eastern side at X-band, while three components show
polarisation at 15\,GHz.  Unexpectedly, components {\bf a} and {\bf
b} show repolarisation having $DP_{14885/23285} = $ 2.6 and 2.2
respectively, while component {\bf c1} is strongly depolarised
($DP_{14885/23285}$=0.14).  Component {\bf a} is the only one found to
be polarised at X-band.  It shows depolarization going from 15 to
8.0\,GHz.

1416$+$067 has been the target of many sub- and milli-arcsecond
resolution observations. It has been observed by \citet{A-G95} at 1.6
and 8.4\,GHz with the VLA, \citet{Ludke98} with MERLIN at 5.0\,GHz, and
by \citet{breugel92} at  15\,GHz with the VLA. We incorporated the
values of $\chi$ measured for some of the components by these observers
to better calculate the RMs. Components {\bf a} and {\bf b} show
RM$_{\it rf}$=--883 and --310 rad m$^{-2}$, respectively. However, we
obtained a much better fit to the $\lambda^2$ law for component {\bf
b} (RM$_{\it rf}$=--1042 rad m$^{-2}$) if the $\chi$ value at
5.0\,GHz of \citet{Ludke98} is discarded.

The structure of 3C298 at milli-arcsecond resolution is discussed in
depth by \citet{Fanti02} and references therein. They observed the
source with the EVN at wavelengths of 92, 18 and 6\,cm.

\noindent
{\bf 1458$+$718 (3C309.1) --}
The sub-arcsecond structure of 1458$+$718 is characterised by several
components. Three of these are aligned in the east-west
direction, while the others lie along the extended emission pointing
south, which is clearly detected in both of our X-band images.
Polarised emission is detected for the two brightest components, 
{\bf a} and {\bf b}, at all four of our observing frequencies

3C309.1 has been the target of many other investigations. We
mention MERLIN observations at 5\,GHz by \citet{Ludke98}, and by
\citet{A-G95} at 1.4, 5, 8.4, and 15\,GHz. The $\chi$ values reported
by \citet{Ludke98}, obtained from observations made at comparable
resolution, were added to our measurements in order to derive the
RMs of three of the components:  RM$_{\it rf}$=--3920 rad m$^{-2}$ for
component {\bf a}, RM$_{\it rf}$=1146 rad m$^{-2}$ for {\bf a1}, and
RM$_{\it rf}$=10185 rad m$^{-2}$ for {\bf d}.  VLBI images for 3C309.1
can be found in \citet{Xu95} at 1.6\,GHz, and in \citet{Kellermann98}
and \citet{Lister05} at 15\,GHz. We note that the milli-arcsec
images show a north-south core-jet structure, while the source
major axis in images at lower resolution is east-west.

\noindent
{\bf 1524$-$136 (OQ172) --}
We only observed this source at 23.2\,GHz since it had already
been observed by \citet{Mantovani94} with the VLA at 6, 4, and
2\,cm.  It has a double structure, and polarised emission is detected
for both of the components at 23.2\,GHz. At lower frequencies the
polarisation is below the detection limits  for the weaker southern
component.  Combining the observations mentioned here we found the
very large RM$_{\it rf}$  of 20\,411 rad m$^{-2}$ for the stronger
northern component, with virtually no depolarisation from 23.2 to
5.0\,GHz.

OQ172 was observed in VLBI by \citet{Udomprasert97}, who find a
component close to the source nucleus with an RM$_{\it rf}$ up to \~
40\,000 rad m$^{-2}$, and by \citet{Mantovani02}, who pointed out that
the source is a quasar with a two-sided jet, which is unusual for this
class of objects.

\noindent
{\bf 1634$+$628, (3C343) --}
This source shows a core-jet structure in the VLA images at X-band.
The jet (or extension) is not detected at 15\,GHz. Polarised emission
is detected for the brightest component, showing very little
depolarisation and a RM$_{\it rf}$=2764 rad m$^{-2}$. Polarimetric VLBA
observations at 5 and 8.4\,GHz for 3C343 were recently published by
\citet{Mantovani10}.  Two bright, compact components, surrounded by
weak, diffuse emission are detected; both components are polarised.
The whole polarised region depolarises little. However, the $\chi$
values present large changes, with the RM$_{\it rf}$ varying
across the source, with indications of values higher than 6\,000 rad
m$^{-2}$.

\noindent
{\bf 1638$+$124 (4C12.60) --}
The source 1638$+$124 shows a double structure at the higher of our
observing frequencies, appearing slightly resolved at 15\,GHz, but
point-like at  X-band. A value for RM$_{\it rf}$=14\,546 rad
m$^{-2}$  is derived for component {\bf a}, which also shows
$DP_{14885/23285}=0.15$.

\noindent
{\bf 1641$+$173 (3C346) --}
The structure of 3C346 is that of an asymmetric triple source.  The
western lobe is completely resolved out at 15 and 23.2\,GHz.  Strong
polarised emission (up to 17\%) is found for the brightest component of
the eastern lobe. This component depolarises quickly with
$DP_{14885/23285}=0.51$, while the RM$_{\it rf}$ is below 73 rad
m$^{-2}$.  This source has also been observed at 1.6, 5, and 8.4\,GHz
by \citet{A-G95} at sub-arcsecond resolution, and by \citet{Spencer91}
with MERLIN and EVN at 18\,cm, and \citet{Cotton95} at 1.7 and 8.4\,GHz

\noindent
{\bf 1829$+$290 (4C29.56) --}
This source possesses a triple structure. The central component is
the brightest and is polarised at both X-band and 15\,GHz, while
polarised emission is not detected at 23.2\,GHz. Weak polarised
emission is also detected for the western lobe at X-band. The two
external lobes and the two-sided jet are below the detection limits at
the two higher frequencies. The derived RM$_{\it rf}$ of 4963 rad
m$^{-2}$ is questionable.

1829$+$290 was observed with the EVN at 1.6\,GHz by
\citet{Dallacasa95}. Their image shows symmetric, bright, elongated
features, likely to be jets, at a position angle of about
90$^\circ$.  This is strongly misaligned with respect to the outer
lobes seen on the sub-arcsecond resolution images. The radio core is
not detected by Dallacasa et al.

\noindent
{\bf 2247$+$140 (4C14.82) --}
The 23.2\,GHz image of this source shows it to have a double
structure.  At the lower resolution of X-band, it appears to be
point-like, while the double structure can be recognised at 15\,GHz.
Although, the structure of the polarised emission allowed us to
easily separate the two regions, even at the lower resolution,
it is difficult to derive the percentage polarisations. It is
easier to derive the $\chi$ values for the two components, and this
has allowed us to calculate RM$_{\it rf}$=319 and 692 rad
m$^{-2}$.

\section{Discussion} 
\label{sec:discussion} 
\subsection{Fractional polarisation} 
Considering the values of the fractional polarisation listed in
Table\,1, our VLA A-array observations confirm that CSSs are weakly
polarised. However, we find that quasars are more highly polarised
than galaxies, and in both cases the polarisation percentages, m,
decrease from higher to lower frequencies. This is consistent with
a number of earlier results (see e.g. Saikia, Swarup \& Kodali 1985;
Saikia, Singal \& Cornwell 1987; Saikia \& Salter 1988, Akujor \& 
Garrington 1995; 
Cotton et al. 2003a). The differences between galaxies and quasars
could be due to a combination of orientation effects and the relative 
contribution of the jet, which is often more strongly polarized. 
Point-like sources are the
least polarised with a median value of m$\simeq$1\%.  The median m
found at X-band confirms the finding by \citet{Mantovani09} that the
median value of m for a complete sample of CSSs observed at the same
band was lower than 1\%.

The median value of m for individual components in both classes of
optical objects decreases slightly going from higher to lower
frequencies.  However, considering galaxies and quasars separately, we
note that the above is confirmed for quasars down to 8085\,MHz, while
galaxies, in contrast, show strong depolarisation between 14885 and
8085\,MHz. 

A comparison can be made with the sample of steep spectrum extended
radio sources selected from the B3-VLA sample observed at 4.85\,GHz by
\citet{klein2003}. These authors found a fractional polarisation
of 5.2\%.

\subsection{Rotation Measures}
RMs have been calculated for point sources and for many individual
components of the sources that are resolved by the VLA A-array.  The
median values for the modulus $\mid$RM$_{\it rf}$$\mid$ are
1146$^{+2183}_{-531}$ rad m$^{-2}$ for the whole set of source 
components,
692$^{+887}_{-532}$ rad m$^{-2}$ for components in quasars, and
1617$^{+2135}_{-1298}$ rad m$^{-2}$ for components in galaxies. We
have 10 galaxies and 6 quasars containing components with RMs
higher than the $\mid$RM$\mid$, while 4 galaxies and 13 quasars have
RMs lower than this.  The dispersion of the RMs is large, ranging
from 0 rad m$^{-2}$ to about 36,000 rad m$^{-2}$ in the source rest
frame.

A narrower dispersion was found for a sample of 47 CSSs observed
with the Effelsberg 100-m telescope by \citet{Mantovani09}.  RMs for
the 16 sources with polarised emission greater than the detection
limits lie between $-20$ and 3900 rad m$^{-2}$. This might be
expected since sources that are unresolved by single dish
observations should show lower levels of polarisation (and RMs)
due to the blending of emission from multiple components having
different EVPAs.  However, among the 12 point-like sources of the
present sample, half have RMs close to zero, while 6 show RMs
$>$500 rad m$^{-2}$, of which 3 are classified as galaxies.

For comparison, for a sample of faint blazars observed with the
Effelsberg 100-m telescope it is found that the RMs range between 0
and 1950 rad m$^{-2}$. Only 9 out of the 27 objects with
polarised emission above the detection limits show RM $>$200 rad
m$^{-2}$ (\citet{Mantovani11}.

The lower values of RM for the components in the galaxies compared
with quasars is interesting. In the unified scheme this could arise
if the quasar components are seen largely through the less dense
ionization cones, while the galaxy components may be seen through the
torus or the denser interstellar medium of the host galaxy. The errors
on the median values are rather large now, and it would be useful to 
extend such studies to a larger sample of sources. Also although 
single-dish values have yielded lower values of RM, it is important
to study these sources with high resolution to determine their
polarimetric properties. A deeper study of the results achieved
by this investigation will be given in a future paper.

\section{Summary and conclusions}
\label{sec:summary}

We have presented multi-frequency VLA polarisation observations of CSSs
and estimated their percentage polarisations and RM values.

About half of the sources are point-like, even at the
$\sim0.1\times0.1$ arcseconds resolution achieved by the present VLA
A-array 23.2-GHz observations. The remaining sources have double or
triple structures. One source, 1210$+$134 (4C13.46) shows a complex
structure.

Low values for the polarisation percentage in CSSs is confirmed by
the present observations. On the average, quasars are more highly
polarised than galaxies.

We have compiled the available $RM$ estimates for CSS sources as seen
on a sub-arcsecond scale. These show a wide range of values, with
indications of very large RMs, although some of the values need to be
confirmed via observations at a larger number of frequencies.  Values
of RM$_{\it rf}$ as high as $\approx$36\,000 rad m$^{-2}$ have been
found.  CSS galaxies are characterized by RM values that are larger
than those found for CSS quasars.  A majority of the objects show
very large values of RM$_{\it rf}$.

\begin{acknowledgements}
The VLA is operated by the U.S. National Radio Astronomy Observatory,
which is a facility of the National Science Foundation operated under a
cooperative agreement by Associated Universities, Inc.  This research
has made use of data from the MOJAVE database that is maintained by the
MOJAVE team (Lister et al., 2009).  It has also
used the NASA/IPAC Extragalactic Database (NED) which is operated by
the Jet Propulsion Laboratory, California Institute of Technology,
under contract with the National Aeronautics and Space Administration,
and NASA's Astrophysics Data System.  We regret that it took so long
to make these data public.  
We are very grateful to the referee, Prof. Ralf Spencer, for the very helpful 
comments and suggestions he made, and for a careful reading of the manuscript 
of this paper.

\end{acknowledgements}

\end{document}